\documentclass{solarphysics}    % Specifies the document style.
\usepackage[optionalrh]{spr-sola-addons}%{solaheader} 
\usepackage{solaheader}
\usepackage{graphicx}
\usepackage{lscape}
\usepackage{longtable}
\newdisplay{guess}{Conjecture}

\usepackage{color}           % For color text: \color command
\usepackage{url}             % For breaking URLs easily trough lines
\begin{document}                                                                                   
\begin{article}
\begin{opening}         
\title{Cross-Calibration of SMART-1 XSM with GOES and RHESSI}
\author{Mikko~\surname{V\"a\"an\"anen\sep} Lauri~\surname{Alha\sep}  Juhani~\surname{Huovelin}}   
\runningauthor{M. V\"a\"an\"anen {\it et al.}}
\runningtitle{Cross-Calibration of SMART-1 XSM with GOES and RHESSI}
\institute{Observatory, P.O. Box 14 FIN-00014 University of Helsinki, Finland email: \url{mikko.vaananen@helsinki.fi} }
\date{submitted: 28 April 2009; accepted: 15 September 2009}

\begin{abstract}
 A number of X-ray instruments have been active in observing the solar coronal X-ray radiation this decade. We have compared {\rm XSM} observations with simultaneous {\rm GOES} and  {\rm RHESSI} observations. We present flux calibrations for all instruments, and compare {\rm XSM} and {\rm GOES} total emission measures (TEM) and temperatures ({\it T}). 

The model-independent flux comparison with {\rm XSM} and {\rm GOES} data at the 1\,--\,8 {\rm \AA}~band shows that the fluxes agree with a ratio of $0.94 \pm 0.09$ for the data up to April 2005. The Mewe model-dependent {\it T}s and TEMs differ as {\rm XSM} observes $1.47 \pm 0.03$ times higher {\it T}s than {\rm GOES} and $1.23 \pm 0.08$ times higher TEMs and $0.92 \pm 0.05$ times lower fluxes. 
The comparison with {\rm RHESSI} data at the 6\,--\,8 {\rm keV} band shows that the average {\rm XSM}/{\rm RHESSI} flux ratio is $2.63 \pm 0.23$.
The discrepancies revealed in this study were similar to discrepancies observed in a number of other spaceborne cross-calibration studies.  

\end{abstract}

\keywords{flares, solar corona, Sun, {\rm SMART-1}, {\rm XSM}, X-rays, cross-calibration, {\rm RHESSI}, {\rm GOES}}

\end{opening}           

\section{Introduction}  
 
Spaceborne instruments have been cross-calibrated in the past, and for
example \citeauthor{Maiz-Apellaniz05} (\citeyear{Maiz-Apellaniz05}) has
cross-calibrated {\it Tycho-2} \cite{Hog00} photometry from {\rm ESA}'s
{\it Hipparcos} and {\it Hubble Space Telescope Spectrophotometry}
\cite{Turnshek90}.  ~In some cases cross-calibration is taken
to mean a calibration with a standard candle, such as the Crab
Nebula, as is the case for {\it International Gamma-Ray Astrophysical
Laboratory} ({\rm INTEGRAL}; \citeauthor{Winkler03},
\citeyear{Winkler03} by \citeauthor{Lubinski04}
(\citeyear{Lubinski04}). Occasionally one also sees the cross-calibration
of different instruments on the same mission, as is the case
for {\it XMM-Newton} \cite{Jansen01} by \citeauthor{Kirsch04}
(\citeyear{Kirsch04}).

The situation was similar for the solar instrument {\it
Solar and Heliospheric Observatory} ({\rm SOHO})  described by
\citeauthor{Domingo95} (\citeyear{Domingo95}) when {\it Coronal
Diagnostic Spectrometer} ({\rm CDS})  and {\it Solar Ultraviolet
Measurements of Emitted Radiation} ({\rm SUMER})  were intercalibrated in
\citeauthor{Pauluhn02} (\citeyear{Pauluhn02}). For the solar instrument
{\it Extreme ultraviolet Imaging Telescope} ({\rm EIT}) and {\rm CDS-NIS}
onboard {\rm SOHO} a sophisticated cross-calibration was recently
done with {\it Transition Region and Coronal Explorer} ({\rm TRACE};
\citeauthor{Handy99}, \citeyear{Handy99}) by \citeauthor{Brooks06} (\citeyear{Brooks06}). In this
cross-calibration the different 171 {\rm \AA}, 195 {\rm \AA}, and 284 {\rm
\AA}~channel fluxes were compared with predicted count rates generated
from a {\it Differential Emission Measure} (DEM) distribution derived from
{\rm CDS} spectral line intensities. The {\rm DEM} was convolved with {\rm
EIT} and {\rm TRACE} temperature response functions, which were calculated
with the latest atomic data from the CHIANTI database \cite{Dere97}
(\url{http://www.ukssdc.ac.uk/solar/chianti/}), \cite{Landi06}, to
predict count rates in their observing channels.

\citeauthor{Stepnik03} (\citeyear{Stepnik03}) presents a cross-calibration where {\it PROgramme National d'AstrOnomie Submillimetrique}
({\rm PRONAOS}; \citeauthor{Serra02}, \citeyear{Serra02}),
a stratospheric balloon-borne submillimetre instrument was cross-calibrated with the {\rm ISOPHOT} photo-polarimeter of \citeauthor{Lemke96} (\citeyear{Lemke96}) onboard {\rm ESA}'s {\it Infrared Space Observatory}
({\rm ISO}; \citeauthor{Kessler96}, \citeyear{Kessler96})
and {\it Diffuse Infrared Background Experiment} {\rm DIRBE}  onboard {\rm NASA}'s {\it Cosmic Background Explorer} {\rm COBE}  described for example in \citeauthor{Boggess92} (\citeyear{Boggess92}).

The current paper presents two sets of cross-calibrations, a model-independent calibration and a Mewe model (\citeauthor{Mewe85} \citeyear{Mewe85}) dependent one. The Mewe model was chosen over the CHIANTI because the Mewe model was provided in all data analysis software systems of all instruments.

{\it Small Missions for Advanced Research and Technology} ({\rm SMART-1};
\citeauthor{Foing03},  \citeyear{Foing03})
was launched on 27 September 2003, and {\it X-ray Solar Monitor} ({\rm XSM};
\citeauthor{Huovelin02},
\citeyear{Huovelin02})
is the only instrument on {\rm SMART-1} for direct observations of the Sun. {\it Reuven Ramaty High Energy Solar Spectroscopic Imager}
({\rm RHESSI}; \citeauthor{Lin02}, \citeyear{Lin02})
is a {\rm NASA} mission launched in 2002 designed to investigate particle acceleration and energy release in solar flares.  {\it Geostationary Operational Environmental Satellite} ({\rm GOES};
\citeauthor{Thomas85}, \citeyear{Thomas85})
is a constellation of weather satellites, where each {\rm GOES} satellite carries also a solar X-ray sensor. The {\rm XSM} spectral range overlaps with {\rm GOES} and {\rm RHESSI}. Concurrent events have been observed and the thrust of this paper is to cross-calibrate the instruments with these events. A further objective of this paper is to develop an understanding of the differences found.

\section{Cross-Calibration Methods}

\subsection{XSM Ground and Inflight Calibrations}

{\rm XSM} itself has been calibrated on the ground. Laboratory calibrations of {\rm XSM} are explained in \citeauthor{Alha08} (\citeyear{Alha08}), and the radiation hardness of {\rm XSM} and the inflight degradation due to space radiation have been studied by \citeauthor{Laukkanen05} (\citeyear{Laukkanen05}). {\rm XSM} is equipped with an inflight spectral calibration source attached to the inner surface of a tungsten shutter. The calibration source consists of $^{55}${\rm Fe} that is coated with a 5 $\mu$m {\rm Ti} foil and produces emission lines at 4.508 and 4.932 {\rm keV} ({\rm Ti}) and 5.895 and 6.492 {\rm keV} ({\rm Mn}). The inflight calibration process is also explained in \citeauthor{Alha08} (\citeyear{Alha08}). 

\subsection{Background Subtraction}

Background subtraction of {\rm XSM} data is done based on measured quiescent Sun background spectra integrated over long time periods.
~Figure 1 shows the quiescent solar spectrum on 6 January 2006 on the left and all sky background spectrum on the right. The flux values derived from these spectra are used in this article when mentioned. Dynamic background subtraction is not possible for all the flares, because not all observations comprise pre- or post flare measurements.

 \begin{figure}
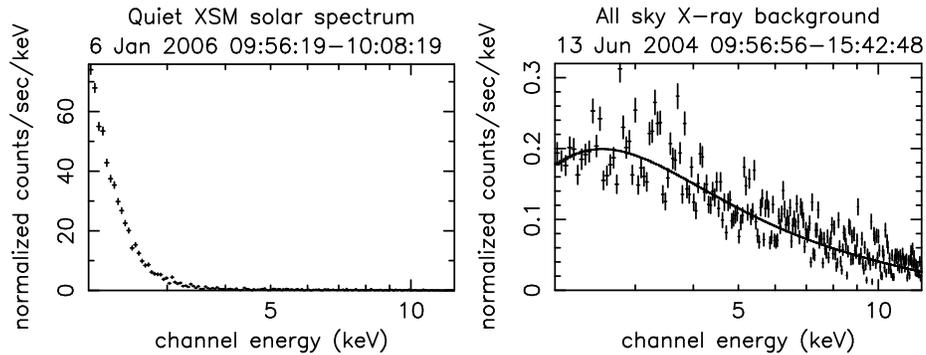
    
   \centerline{\hspace*{-0.0\textwidth}
               \includegraphics[width=0.38\textwidth,angle=270]{fig1a.ps}
               \hspace*{0.0\textwidth}
               \includegraphics[width=0.38\textwidth,angle=270]{fig1b.ps}
              }
     \vspace{0.0\textwidth}   
      \caption[]{The {\rm XSM} quiescent solar spectrum on 6 January 2006 09:56:19\,--\,10:08:19 on the left. The non-solar all sky background on 13 June 2004 09:56:56\,--\,15:42:48 on the right. The line in the right plot is the fitted cutoff powerlaw. }
\label{penG}

      \end{figure}

The background flux values deduced from the quiescent solar spectrum were $2.7\times10^{-8}$ {\rm W m}$^{-2}$ for 1.55\,--\,12.40 {\rm keV} and $1.0\times10^{-10}$ {\rm W m}$^{-2}$ for 6\,--\,8 {\rm keV}. The all sky background flux values deduced from the spectrum were $1.0\times10^{-9}$ {\rm W m}$^{-2}$ for 1.55\,--\,12.40 {\rm keV} and $1.7\times10^{-10}$ {\rm W m}$^{-2}$  for 6\,--\,8 {\rm keV} fitted with a cutoff powerlaw between 2\,--\,12.4 {\rm keV}.

\subsection{{\rm XSM} and {\rm GOES} Model-Independent Calibrations}

 The {\rm XSM} flux {\it F}$_{{\rm XSM}}$ was derived from XSPEC \cite{Arnaud96} (\url{http://heasarc.gsfc.nasa.gov/docs/xanadu/xspec/}) spectral fits that sampled the data best. We used one-minute {\rm GOES} data {\it F}$_{{\rm GOES}}$ and 16-second {\rm XSM} data to derive the flux values listed in Table 1. We interpolated {\rm XSM} measurements to match with the {\rm GOES} measurements in time with one second accuracy. 
 
 We are quoting the $\sigma$ error derived in this way for the errors, unless otherwise stated. 

\subsection{{\rm XSM} and {\rm GOES} Mewe-Model Calibrations}

In addition to the actual flux calibration we obtained the
{\rm GOES} total emission measures (TEM), TEM$_{{\rm GOES}}$ and
temperatures ({\it T}), {\it T}$_{{\rm GOES}}$ using the Mewe model
of \citeauthor{Mewe85} (\citeyear{Mewe85}) with Meyer abundances
\cite{Meyer85} from the {\rm GOES} routine in \url{SolarSoft} \cite{Freeland98}
(\url{http://www.lmsal.com/solarsoft/index_old.html}). We fitted the Mewe
model to {\rm XSM} data in XSPEC using the ``mekal'' algorithm from 2.0
{\rm keV} onwards to obtain the corresponding TEM$_{{\rm XSM}}$, {\it
T}$_{{\rm XSM}}$ and {\it F}${\rm_{XSM}^{Mewe}}$ values. The band between
1.55\,--\,2.0 {\rm keV} needs to be extrapolated due to limitations
cited in \citeauthor{Alha08} (\citeyear{Alha08}). We then compared
the Mewe model-generated fluxes {\it F}${\rm_{XSM}^{Mewe}}$ and {\it
F}${\rm_{GOES}^{\rm Mewe}}$, which we obtain by feeding the {\rm GOES}
routine values from Solarsoft to ``mekal'' in XSPEC through the {\rm XSM}
response. {\rm TEM} and {\it T} refer to the Mewe generated values for
both instruments in {\rm XSM}-{\rm GOES} calibrations of this paper. Mewe
generated fluxes ({\it F}) are mentioned explicitly for both instruments.

The results of these calibrations are discussed in Sections 3.1 and 3.2 and the first conclusion in Section 5.

\subsection{{\rm XSM} and {\rm RHESSI}}

{\rm XSM} and {\rm RHESSI} were cross-calibrated in the 6\,--\,8 {\rm keV} band. This is the band where the sensitivities of the two instruments are most similar (B. Dennis, H. Hudson, private communication, 7-11 Jun 2005). 

The dynamic pre- and postflare background was subtracted for {\rm RHESSI}. The quiescent {\rm XSM} background was subtracted for {\rm XSM}. {\rm XSM} fluxes were calculated by fitting the vRaymond \cite{Raymond77} + broken powerlaw model between 5\,--\,10 {\rm keV} in XSPEC and ``vth'' using the Mewe full model in OSPEX (\url{http://hesperia.gsfc.nasa.gov/ssw/packages/spex/doc/ospex_explanation.htm}) of Solarsoft \cite{Freeland98} was used to derive {\rm RHESSI} fluxes. {\rm XSM} data was also fed into OSPEX, and the two models produced the same flux results independently in both XSPEC and OSPEX. Therefore any differences in software or model methodology are ruled out as sources of discrepancy.

\section{Cross-Calibration Results}

In the following we describe the cross-calibration results obtained from each pair of instruments individually.

\subsection{{\rm XSM} and {\rm GOES} Model-Independent Calibrations}

The light curves of Figure 2 demonstrate that {\rm XSM} and {\rm GOES} are working coherently in time.

 \begin{figure}    
    \vspace{0.0\textwidth} 
    \hspace*{-0.05\textwidth}
   \centerline{\includegraphics[width=0.7\textwidth,angle=0]{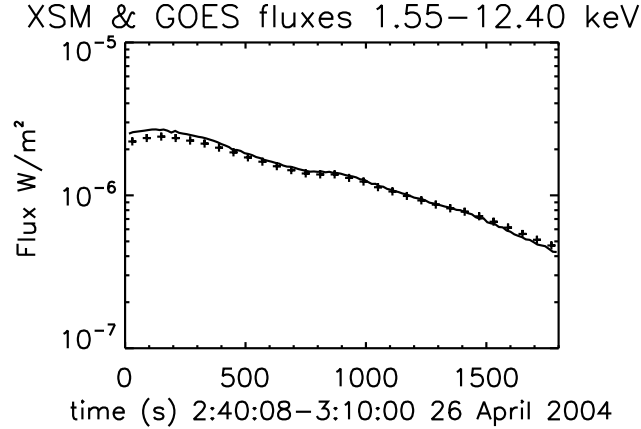}}
     \vspace{-0.03\textwidth} 
     \caption[]{The plot shows both {\rm GOES} (+signs) 1\,--\,8 {\rm \AA}~and {\rm XSM} (solid line) light curves on the 26 April 2004 2:40:08\,--\,3:10:00.}
\label{penG}
      \end{figure}

Table 1 provides a comprehensive list of all model-independent {\rm XSM}-{\rm GOES} flux calibrations performed in chronological order. Table 1 also displays the {\rm XSM}/{\rm GOES} flux ratios and their errors.  The flux ratio is between 1.23 and 0.69. The average flux ratio is $0.94 \pm 0.09$. {\it $\theta^{\rm off-axis}_{{\rm XSM}}$} is the angle between the Sun and the optical axis of {\rm XSM}.

\begin{table}
\tabcolsep=3pt
\centering
\caption[]{Comprehensive calibration list of fluxes of {\rm GOES} and {\rm XSM} measured in the 1.55\,--\,12.40 {\rm keV} band during one-minute periods in chronological order from April 2004 to April 2005. Errors are the standard deviations of the mean $\sigma$. }
\label{sphericcase}
\begin{tabular}{crrrrrrrc}
\hline
\\
Interval & \multicolumn{1}{c}{time} & \multicolumn{1}{c}{{\it F}$_{{\rm GOES}}$} & \multicolumn{1}{c}{{\it F}$_{{\rm XSM}}$} & \multicolumn{1}{c}{{\it F}$_{{\rm XSM}}$/{\it F}$_{{\rm GOES}}$} & \multicolumn{1}{c}{{$\theta$}$^{\rm off-axis}_{{\rm XSM}}$} &\\
 & \multicolumn{1}{c}{ ({\rm min}) } & \multicolumn{1}{c}{({ \rm W m}$^{-2}$)} & \multicolumn{1}{c}{ ({\rm W m}$^{-2}$)} & \multicolumn{1}{c}{} & \multicolumn{1}{c}{ ({\rm deg})} &\\

\hline

$1$ & 03:02 26 Apr 2004 & $8.22\times10^{-7}$ &$(8.28 \pm 0.37)\times10^{-7}$& $(1.01 \pm 0.04)$&$4$\\
$2$ & 19:38 5 May 2004  & $8.38\times10^{-8}$ & $(7.34 \pm 0.86)\times10^{-8}$ & $(0.88 \pm 0.12)$ & $13$ \\
$3$ & 7:30 24 May 2004  & $3.28\times10^{-7}$ & $(2.90 \pm 0.89)\times10^{-7}$ & $(0.88 \pm 0.31)$ & $39$ \\
$4$ & 11:06 24 May 2004 & $4.47\times10^{-6}$ & $(4.57 \pm 1.16)\times10^{-6}$ & $(1.02 \pm 0.25)$ & $39$ \\
$5$ & 05:47 16 Jun 2004 & $5.65\times10^{-7}$ & $(3.91 \pm 1.05)\times10^{-7}$ & $(0.69 \pm 0.27)$ & $41$ \\
$6$ & 20:27 5 Jul 2004  & $3.43\times10^{-8}$ & $(3.29 \pm 0.79)\times10^{-8}$ & $(0.96 \pm 0.24)$ & $14$ \\
$7$ & 05:31 31 Jul 2004 & $4.14\times10^{-7}$ & $(3.93 \pm 0.02)\times10^{-7}$ & $(0.95 \pm 0.01)$ & $17$ \\
$8$ & 05:27 25 Aug 2004 & $7.01\times10^{-7}$ & $(6.19 \pm 0.64)\times10^{-7}$ & $(0.88 \pm 0.10)$ & $12$ \\  
$9$ & 05:59 15 Jan 2005 & $7.34\times10^{-6}$ & $(9.00 \pm 1.13)\times10^{-6}$ & $(1.23 \pm 0.13)$ & $16$ \\
$10$ & 12:51 4 Apr 2005 & $2.90\times10^{-7}$ & $(2.71 \pm 0.10)\times10^{-7}$ & $(0.93 \pm 0.04)$ & $26$ \\
\\ \hline
\vspace*{-0.1\textwidth}
\end{tabular}
\end{table}

In the top plot of Figure 3 the {\it F}$_{{\rm XSM}}$/{\it F}$_{{\rm GOES}}$ ratio is plotted as a function of {\it F}$_{{\rm XSM}}$. There appears to be no significant trend in this ratio with {\it F}$_{{\rm XSM}}$. If interval 6, an essentially quiescent interval is omitted, the {\it F}$_{{\rm XSM}}$/{\it F}$_{{\rm GOES}}$ ratio is also $0.94 \pm 0.09$.

  \begin{figure}   
    \vspace{0.0\textwidth} 
   \centerline{\hspace*{-0.05\textwidth}
               \includegraphics[width=0.5\textwidth]{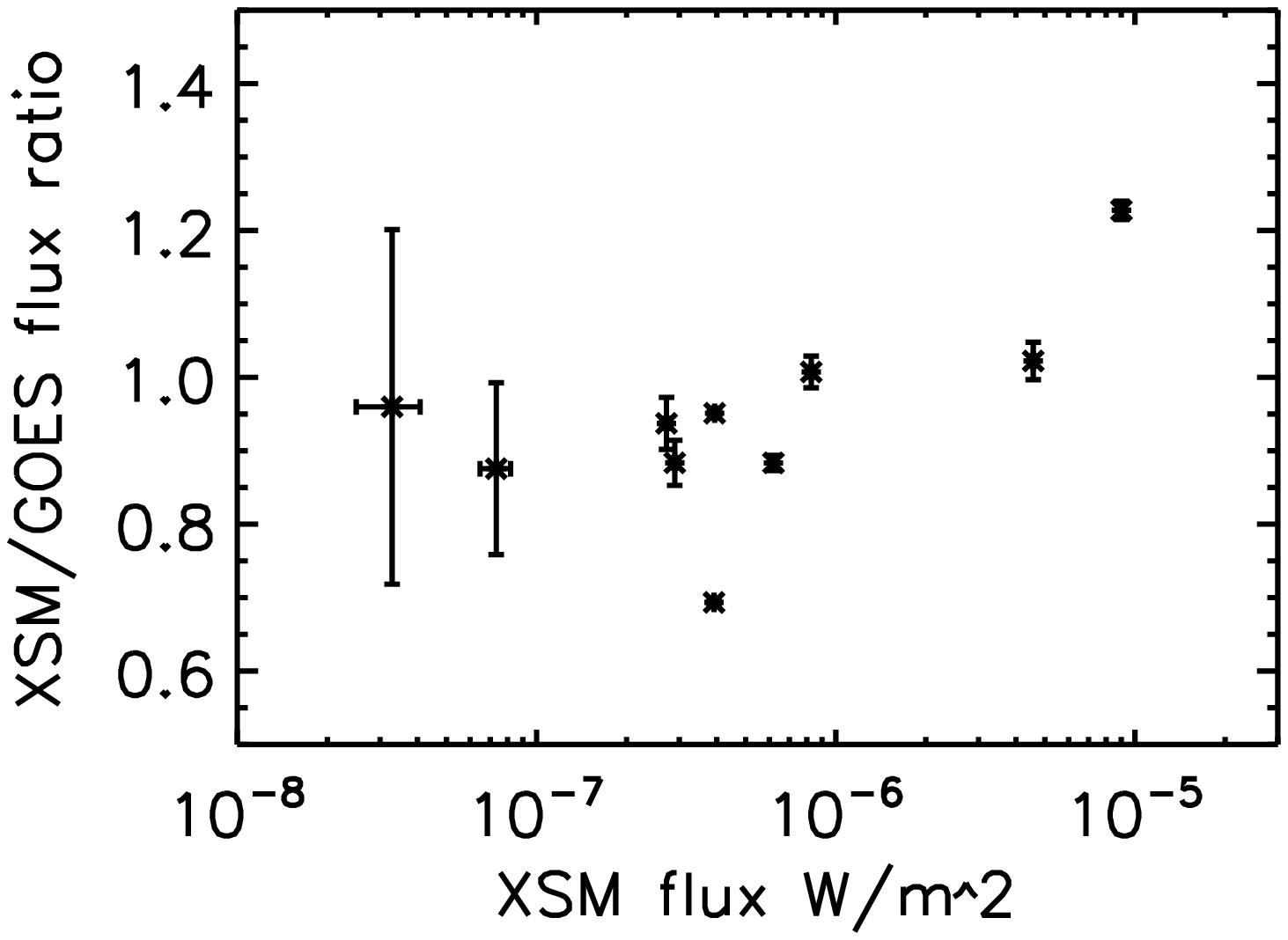}
               \hspace*{-0.05\textwidth}
               \includegraphics[width=0.5\textwidth]{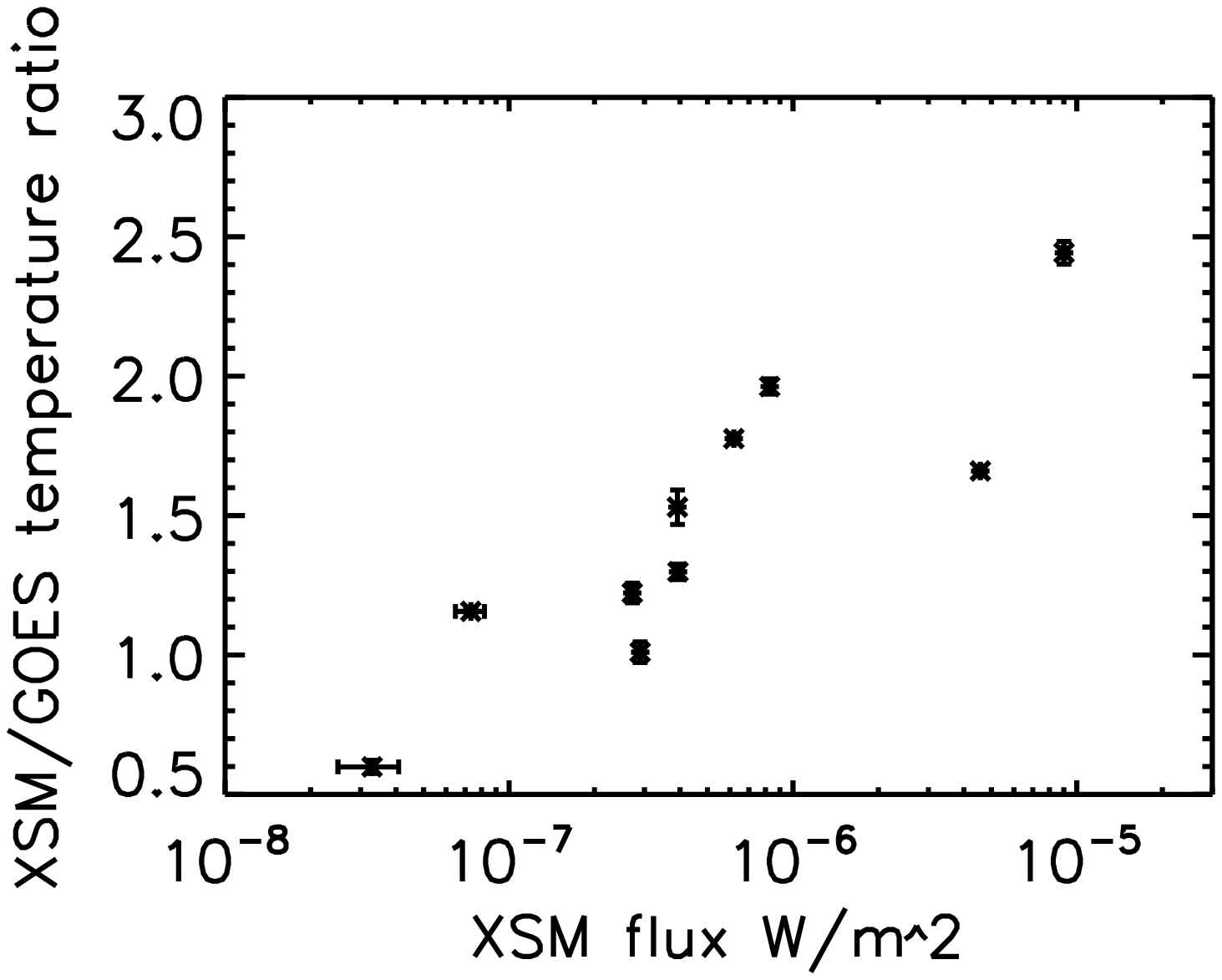}
              }
     \vspace{-0.03\textwidth}   
        \centerline{\hspace*{-0.05\textwidth}
               \includegraphics[width=0.5\textwidth]{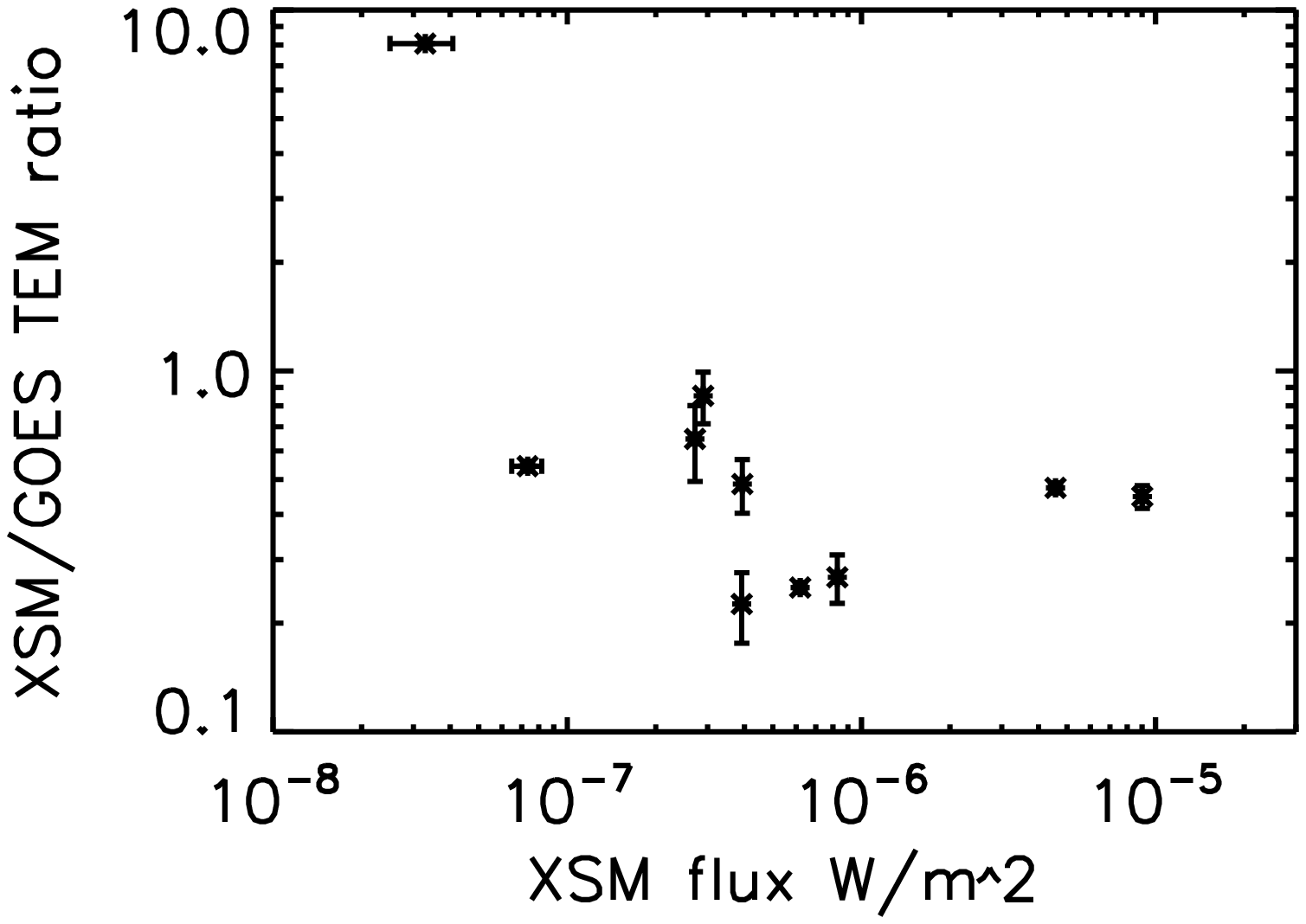}
               \hspace*{-0.05\textwidth}
               \includegraphics[width=0.5\textwidth]{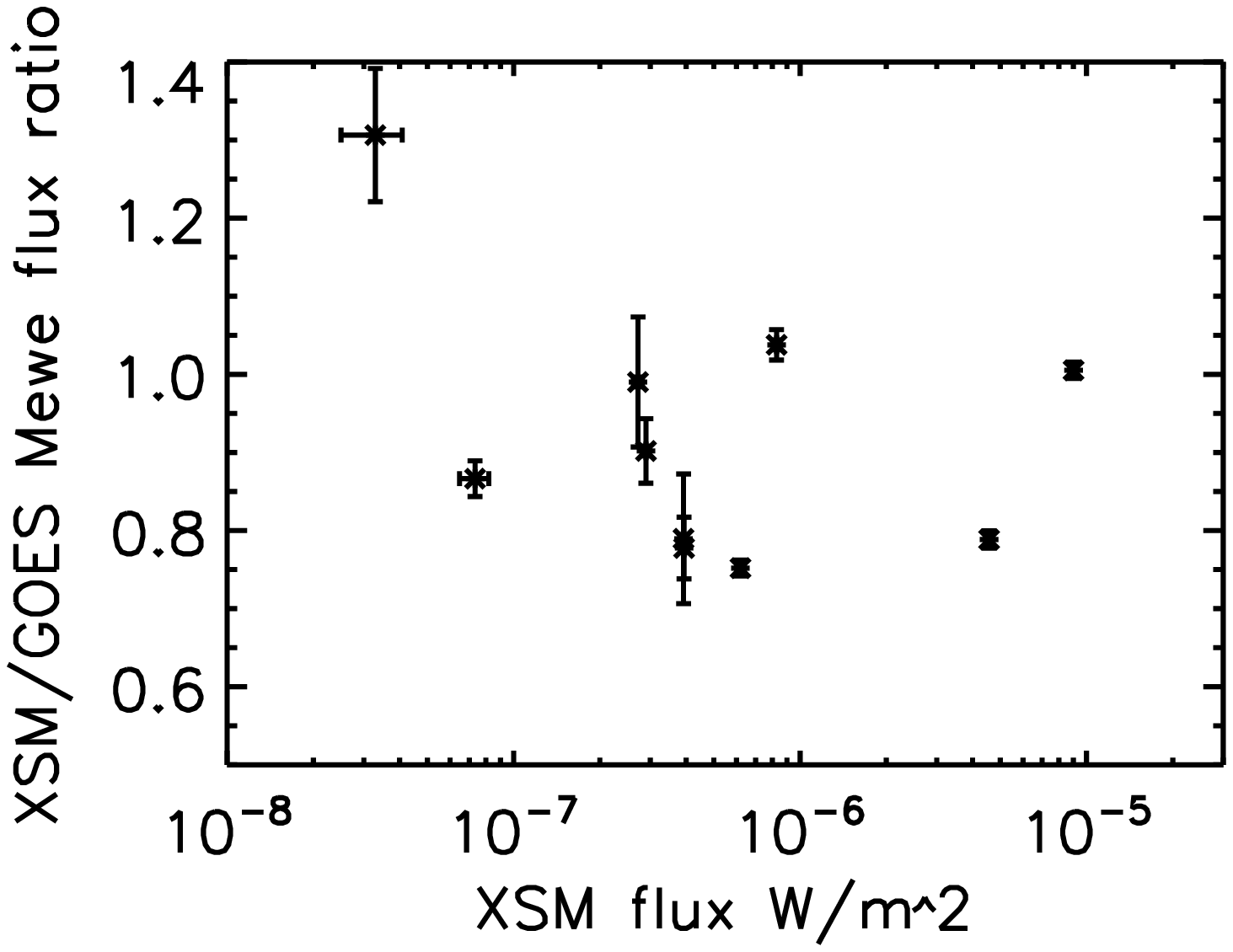}
              }  
\vspace{-0.02\textwidth}
 \caption[]{{\it F}$_{{\rm XSM}}$/{\it F}$_{{\rm GOES}}$ ratios plotted as a function of {\rm XSM} flux (Table 1) on top left. On top right the {\it T}$_{{\rm XSM}}$/{\it T}$_{{\rm GOES}}$ plotted as a function of {\rm XSM} flux. On bottom left the TEM$_{{\rm XSM}}$/TEM$_{{\rm GOES}}$ is plotted as a function of {\rm XSM} flux. On bottom right {\it F}${\rm _{XSM}^{Mewe}}$/{\it F}${\rm _{GOES}^{Mewe}}$ is plotted as a function of {\rm XSM} flux. Interval 6 is the lowest point in top right plot and highest point in lower-left plot.}
\label{penG}             
   \end{figure}
\vspace{-5pt}

\subsection{{\rm XSM} and {\rm GOES} Mewe-Model calibrations}

Table 2 presents the {\rm GOES} and {\rm XSM} fluxes ({\it F}${\rm_{GOES}^{Mewe}}$, {\it F}${\rm_{XSM}^{Mewe}}$), TEMs (TEM$_{{\rm GOES}}$, TEM$_{{\rm XSM}}$) and temperatures ({\it T}$_{{\rm GOES}}$, {\it T}$_{{\rm XSM}}$) obtained from the Mewe model with Meyer abundances.  The average {\it F}${\rm_{XSM}^{Mewe}}$/{\it F}${\rm_{GOES}^{Mewe}}$ ratio is $0.92 \pm 0.05$, meaning that the {\rm GOES} response produces the same flux with the Mewe model in comparison to {\rm XSM}. The average {\rm TEM}$_{{\rm XSM}}$/{\rm TEM}$_{{\rm GOES}}$ ratio is $1.23 \pm 0.08$. The {\rm XSM} temperatures fitted with the Mewe model are about 50\% higher; the average {\it T}$_{{\rm XSM}}$/{\it T}$_{{\rm GOES}}$ ratio equals $1.47 \pm 0.03$. 

As we can see from Table 2 and Figure 3, interval 6 deviates quite far from the general trend in Figure 3. This is because it is essentially a quiescent interval. In this quiet state the {\it F}${\rm_{XSM}^{Mewe}}$/{\it F}${\rm_{GOES}^{Mewe}}$ ratio is $1.30 \pm 0.09$, {\it T}$_{{\rm XSM}}$/{\it T}$_{{\rm GOES}}$ ratio is $0.60 \pm 0.02$ and the {\rm TEM}$_{{\rm XSM}}$/{\rm TEM}$_{{\rm GOES}}$ ratio is $8.07 \pm 0.1$.

\begin{table}
%\vspace*{-0.05\textwidth}
\tabcolsep=3pt
\centering
\caption[]{Events of Table 1 calibrated using the \citeauthor{Mewe85} (\citeyear{Mewe85}) model with \citeauthor{Meyer85} (\citeyear{Meyer85}) abundances for both {\rm XSM} and {\rm GOES} data.}
\label{sphericcase}
\begin{tabular}{crrrrrrrc}
\hline
\\
Interval & \multicolumn{1}{c}{{\it T}$_{{\rm GOES}}$ }& \multicolumn{1}{c}{{\it T}$_{\rm {XSM}}$} & \multicolumn{1}{c}{{\rm TEM}$_{{\rm GOES}}$} & \multicolumn{1}{c}{{\rm TEM}$_{{\rm XSM}}$}& \multicolumn{1}{c}{{\it F}${\rm_{GOES}^{Mewe}}$}& \multicolumn{1}{c}{{\it F}${\rm_{XSM}^{Mewe}}$}\\
 & \multicolumn{1}{c}{({\rm keV})}& \multicolumn{1}{c}{({\rm keV})} & \multicolumn{1}{c}{($10^{47}${\rm cm}$^{-3}$)} & \multicolumn{1}{c}{($10^{47}${\rm cm}$^{-3}$)}& \multicolumn{1}{c}{({\rm W m}$^{-2}$)}& \multicolumn{1}{c}{({\rm W m}$^{-2}$)}\\
\hline

$~~1$ &  $0.48$ & $0.94 \pm 0.01$ & $40$  & $10.8 \pm 0.2$  & $8.83\times10^{-7}$ & $(9.20 \pm 0.14)\times10^{-7}$ \\
$~~2$ &  $0.37$ & $0.42 \pm 0.01$ & $13$  & $7.1 \pm 0.1$  & $1.23\times10^{-7}$ & $(1.07 \pm 0.03)\times10^{-7}$ \\
$~~3$ &  $0.52$ & $0.53 \pm 0.02$ & $17$  & $14.4 \pm 2.0$  & $4.68\times10^{-7}$ & $(4.22 \pm 0.17)\times10^{-7}$ \\
$~~4$ &  $0.94$ & $1.56 \pm 0.02$ & $75$  & $35.6 \pm 0.5$ & $6.39\times10^{-6}$ & $(5.04 \pm 0.06)\times10^{-6}$ \\
$~~5$ &  $0.39$ & $0.60 \pm 0.04$ & $65$  & $14.6 \pm 2.8$ & $7.40\times10^{-7}$ & $(5.84 \pm 0.49)\times10^{-7}$ \\
$~~6$ &  $0.47$ & $0.28 \pm 0.01$ & $2.5$ & $20.1 \pm 2.1$ & $5.19\times10^{-8}$ & $(6.78 \pm 0.58)\times10^{-8}$ \\
$~~7$ &  $0.62$ & $0.80 \pm 0.02$ & $15$  & $7.2 \pm 0.3$ & $6.33\times10^{-7}$ & $(4.92 \pm 0.19)\times10^{-7}$ \\
$~~8$ &  $0.53$ & $0.95 \pm 0.01$ & $35$  & $8.8 \pm 0.1$ & $1.01\times10^{-6}$ & $(7.59 \pm 0.08)\times10^{-7}$ \\  
$~~9$ &  $1.27$ & $3.10 \pm 0.13$ & $80$  & $35.8 \pm 1.2$  & $9.32\times10^{-6}$ & $(9.36 \pm 0.27)\times10^{-6}$ \\
$~~10$&  $0.60$ & $0.74 \pm 0.03$ & $10$  & $6.5 \pm 0.6$ & $3.92\times10^{-7}$ & $(3.88 \pm 0.32)\times10^{-7}$ \\
\\ \hline
\vspace*{-0.05\textwidth}
\end{tabular}
\end{table}

If the quiescent interval 6 is discounted from the averages to obtain pure ``flare-on'' values, the average {\it F}${\rm_{XSM}^{Mewe}}$/{\it F}${\rm_{GOES}^{Mewe}}$ ratio is $0.88 \pm 0.05$, the average {\it T}$_{{\rm XSM}}$/{\it T}$_{{\rm GOES}}$ ratio is $1.56 \pm 0.03$ and the average {\rm TEM}$_{{\rm XSM}}$/{\rm TEM}$_{{\rm GOES}}$ ratio is $0.46 \pm 0.08$. 

Figure 3 shows that the {\it T}$_{{\rm XSM}}$/{\it T}$_{{\rm GOES}}$ increases as a function of {\it F}$_{{\rm XSM}}$. As for the bottom plots, neither {\it F}${\rm _{XSM}^{Mewe}}$/{\it F}${\rm _{GOES}^{Mewe}}$ or {\rm TEM}$_{{\rm XSM}}$/{\rm TEM}$_{{\rm GOES}}$ seem to vary with {\it F}$_{{\rm XSM}}$. 

It should also be noted that towards the higher energy flares the observed spectrum deviates more from the Mewe model. In order to visualise the situation, the Mewe models predicted by {\rm GOES} and {\rm XSM} are plotted against {\rm XSM} data for interval 9, the biggest flare, in Figure 4. The spectral model could be improved with the addition of a high-energy component.

 \begin{figure}
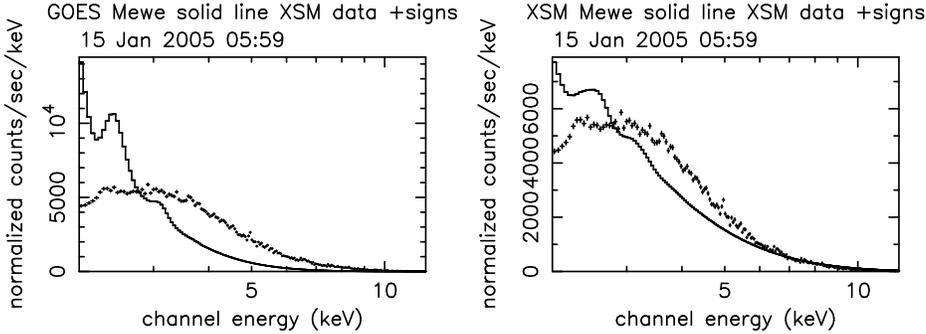
    
   \centerline{\hspace*{-0.0\textwidth}
               \includegraphics[width=0.36\textwidth,angle=270]{fig4a.ps}
               \hspace*{0.0\textwidth}
               \includegraphics[width=0.36\textwidth,angle=270]{fig4b.ps}
              }
     \vspace{0.0\textwidth}   
\caption[]{The Mewe emission spectra predicted by {\rm GOES} (left) and {\rm XSM} (right) is plotted as a solid line, {\rm XSM} data marked by + signs. The event is the largest flare, interval number 9.}
\label{penG}
      \end{figure}

\subsection{{\rm XSM} and {\rm RHESSI}}

Figure 5 displays a longer duration light curve from the decay phase of the same flare as in Figure 1. 

 \begin{figure}    
     \vspace{0.0\textwidth} 
    \hspace*{-0.05\textwidth}
   \centerline{\includegraphics[width=0.7\textwidth,angle=0]{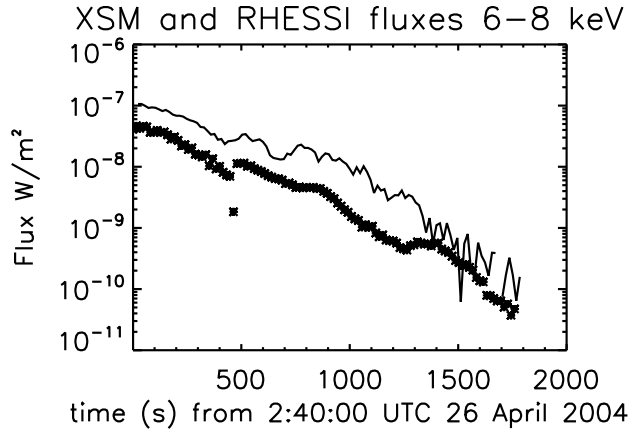}}
    \vspace{-0.02\textwidth} 
    \caption[]{{\rm RHESSI} and {\rm XSM} fluxes plotted as a function of time in the 6\,--\,8 {\rm keV} band. {\rm XSM} flux is the solid line, {\rm RHESSI} flux is marked by crosses.}
\label{penG}
      \end{figure}

The average flux ratio of {\rm XSM} flux/{\rm RHESSI} flux between 6\,--\,8 {\rm keV} was 2.63. Similar measurement errors as in the previous section put the ratio at $2.63 \pm 0.23$, assuming {\rm XSM} errors only. At lower flux levels, the measurements approach each other. The flux ratio is steady around the average at the beginning of the interval, but varies quite randomly between 0.6 to 10 at the end of the measurement interval.

\section{Discussion}

The flux differences between {\rm GOES} and {\rm XSM} appear to be within the measurement error. Half of the calibrations have {\it F}$_{{\rm GOES}}$ and  {\it F}$_{{\rm XSM}}$ within $\sigma$, and 9/10 intervals are within 3 $\sigma$. 
When the Mewe model and Meyer abundances were used with both {\rm XSM} and {\rm GOES} data, {\it T}$_{{\rm XSM}}$ was $1.47 \pm 0.03$ times higher than{\it T}$_{{\rm GOES}}$. In contrast, {\it F}${\rm _{XSM}^{Mewe}}$ was $0.92 \pm 0.05$ times lower than {\it F}${\rm _{GOES}^{Mewe}}$ and {\rm TEM}$_{{\rm XSM}}$ was $1.23 \pm 0.08$ times higher than {\rm TEM}$_{{\rm GOES}}$. We believe that the likely cause for the discrepancy between {\rm XSM} and {\rm GOES} in the Mewe model derived parameters relates to three factors:

{\it i}) The statistics of the data: {\rm GOES} has only two, whereas {\rm XSM} has 512 channels. 

{\it ii}) Extrapolation of {\rm XSM} data between 1.55\,--\,2.0 {\rm keV} from a fit between 2.0\,--\,12.4 {\rm keV} to overcome the practical low energy limitations of {\rm XSM} as explained in \citeauthor{Alha08} (\citeyear{Alha08}).

{\it iii}) The need for a high-energy spectral component.  

The Mewe model is a thermal line emission + continuum model. Figure 4 clearly shows that this model is not as appropriate for estimating the flux, {\it T} or {\rm TEM} with bigger flares, as may the case be with quiescent solar observations or small flares. The Mewe model misses an important part of the high energy flux, which probably has a non-thermal origin. In order to improve upon the predictability of model parameters from {\rm GOES} data it is probably not enough to update the line emission model only, to say CHIANTI for example, as has been done in OSPEX. The {\rm GOES} differences in temperature and emission measure responses observed with different models of Mewe and CHIANTI in \citeauthor{White05} (\citeyear{White05}) are about 25\%, and would suggest that a change in the emission model might compensate for some discrepancies.  Based on the observations made here, the calibration should be repeated with CHIANTI and {\rm XSM} data in the future. 

The {\rm XSM}/{\rm RHESSI} flux ratio is $2.63 \pm 0.23$, where the error derives solely from the estimated error for {\rm XSM}. In order to bring the measurements to within 3 $\sigma$ of each other the relative {\rm RHESSI} flux error should be $\sigma$=0.33. This $\sigma$ may be possible, but in addition there could be systematic effects that amount to the discrepancy observed. Firstly, it should be noted that between 5\,--\,10 {\rm keV} the effective area of {\rm RHESSI} falls over two orders of magnitude as noted in \citeauthor{Smith02} (\citeyear{Smith02}), so defining the effective area is difficult. In this same band the {\rm XSM} effective area varies by less than 5 \% as explained in \citeauthor{Huovelin02} (\citeyear{Huovelin02}). 

During the calibration interval at approximately 26 April 2004 02:48 {\rm RHESSI} changes from A1 state (=thin attenuator on) to A0 state where no attenuators are on. It is probably this event that causes the one single dropped data point in Figure 5 at approximately 480 seconds. However, considering that the {\rm XSM}/{\rm RHESSI} flux ratio behaves normally on both sides of this point, the attenuators are not likely to distort recorded fluxes. 

{\rm XSM} saw a higher photon flux than {\rm RHESSI} at the higher energies. Time integrated average spectra of this interval revealed that the flux was $2.6\times10^{-8}$ {\rm W m}$^{-2}$ between 6.0\,--\,8.0 {\rm keV} averaged over the entire observation period. This is two orders of magnitude higher than quiescent background or all sky background (both about $1\times10^{-10}$ {\rm W m}$^{-2}$) during a {\rm GOES} B-class flare. Therefore there is reason to believe this flux is solar in origin.
 
In order to put these calibration results into perspective they could perhaps be compared with \citeauthor{Brooks06} (\citeyear{Brooks06}) where a discrepancy of 3\,--\,25\% was observed when {\rm CDS} DEMs were used to predict {\rm TRACE} and {\rm EIT} 171 {\rm \AA}~and 195 {\rm \AA}~count rates and the two-to-five fold discrepancy was observed for the 284 {\rm \AA}~count rate. In \citeauthor{Stepnik03} (\citeyear{Stepnik03}) a 0.7 conversion coefficient was obtained between {\rm ISOPHOT} and {\rm PRONAOS}. For the {\rm Ne {\small VIII}} narrow line observation at 77.0 {\rm nm} \cite{Pauluhn02} reported an average ratio of 2.6 for the {\rm CDS-GIS-4} to {\rm SUMER} radiances, when {\rm CDS} measured 30\% higher values than {\rm SUMER} for the {\rm He {\small I}} line at 58.4 {\rm nm}. Remarkably the narrow band calibrations conducted between {\rm XSM} and {\rm RHESSI} show discrepancies similar to those observed by \citeauthor{Pauluhn02} (\citeyear{Pauluhn02}) for {\rm SUMER} and {\rm CDS-GIS-4} detector, or the 284 {\rm \AA}~channel of {\rm TRACE} and {\rm EIT} in \citeauthor{Brooks06} (\citeyear{Brooks06}). The flux calibration between {\rm XSM} and {\rm GOES} shows discrepancies that are smaller or similar to the discrepancies observed in the cross-calibrations of \citeauthor{Brooks06} (\citeyear{Brooks06}) for the other channels and \citeauthor{Stepnik03} (\citeyear{Stepnik03}) and \citeauthor{Pauluhn02} (\citeyear{Pauluhn02}) for the {\rm He {\small I}} line.

\section{Conclusions}

The main conclusions reached in these cross-calibrations were:

{\it i}) The model independent {\it F}$_{ {\rm XSM}}$/{\it F}$_{ {\rm GOES}}$ ratio is $0.94 \pm 0.09$ for data prior to April 2005. {\rm XSM} and {\rm GOES} agree in terms of model independent and Mewe model dependent fluxes. However, discrepancies arise in the model parameters {\it T} and {\rm TEM} predicted by the Mewe model with Meyer abundances. It is suggested that the discrepancies arise from three factors, first of which is the lack of sampling due to the {\rm GOES} data having only two channels in contrast to 512 channels for {\rm XSM}. The second is the extrapolation of the model between 1.55\,--\,2.0 {\rm keV} in the {\it F}${\rm _{XSM}^{Mewe}}$. The third suggested source for discrepancy is an additional high-energy component in the spectral model. 

{\it ii}) The average {\rm XSM}/{\rm RHESSI} flux ratio is $2.63 \pm 0.23$. There are a number of possible sources for discrepancy, one of which is that within the calibration band of 6\,--\,8 {\rm keV} an asymptotic change in {\rm RHESSI} effective area introduces error. 

{\it iii}) The calibration results discovered here are similar to results obtained from other spaceborne cross-calibrations from \citeauthor{Brooks06} (\citeyear{Brooks06}), \citeauthor{Pauluhn02} (\citeyear{Pauluhn02}) and \citeauthor{Stepnik03} (\citeyear{Stepnik03}).

\acknowledgements

This work was made possible by {\rm ESA} staff and other instrument teams involved in the {\rm SMART-1} mission. We also thank Dr Bernard Foing at {\rm ESA} and Dr Manuel Grande in the UK. Acknowledgements are due for the entire {\rm RHESSI} -team for their assistance. In particular we extend our thanks to Drs Brian Dennis, Richard Schwartz,  Kim Tolbert GSFC/NASA and Hugh Hudson, Gordon Hurford, Robert Lin and Jim McTiernan of Space Sciences Lab at University of California at Berkeley and Ken Phillips at Mullard Space Sciences Laboratory, UK. Dr. Jukka Nevalainen is acknowledged for advice that improved the manuscript. This work was funded by the following grants: V\"ais\"al\"a foundation PhD grants, Academy of Finland research grants 74882 and 211061 and Suinno Oy. The anonymous referee is acknowledged for suggesting several improvements and clarifying the paper.

\theendnotes

\end{article}
\end{document}